\documentclass{article}

 \UseRawInputEncoding 

\newtheorem{theorem}{Theorem}[section]
\newtheorem{lemma}[theorem]{Lemma}
\newtheorem{corollary}[theorem]{Corollary}
\newtheorem{definition}[theorem]{Definition}
\newtheorem{problem}[theorem]{Problem}
\newtheorem{conjecture}[theorem]{Conjecture}

\newcommand\qed{\begin{flushright} {\bf q.e.d.} \end{flushright} }
\newcommand\prf{\noindent {\bf Proof :}}  

\newcommand\bits{\{0,1\}}
\newcommand\uu{{\bits^*}}
\newcommand\nn{{\bits^n}}
\newcommand\mm{{\bits^m}}
\newcommand\kk{{\bits^k}}

\newcommand\tru{{\mbox{{\bf tt}}}}
\newcommand\gadf{{\mbox{Gad}_{f}}}
\newcommand\gad{{\mbox{Gad}_{sq}}}
\newcommand\gadc{{\mbox{Gad}^c_{sq}}}
\newcommand\tpv{{T_{\mbox{\tiny PV}}}}
\newcommand\bt{S^1_2 + \mbox{dWPHP}(\Delta^b_1)}

\newcommand\pp{{\cal P}}
\newcommand\np{{\cal N}{\cal P}}
\newcommand\ee{{\cal E}}
\newcommand\nee{{\cal N}{\cal E}}
\newcommand\eee{{{\cal E}{\cal X}{\cal P}}}

\begin{document}

\title{On the existence of strong proof complexity generators}  

\author{Jan Kraj\'{\i}\v{c}ek}

\date{Faculty of Mathematics and Physics\\
Charles University\thanks{
Sokolovsk\' a 83, Prague, 186 75,
The Czech Republic,
{\tt krajicek@karlin.mff.cuni.cz}}}

\maketitle

\begin{abstract}
Cook and Reckhow \cite{CooRec} pointed out that $\np \neq co\np$ iff there is no propositional proof system that admits polynomial size proofs of all tautologies. The theory of proof complexity
generators aims at constructing sets of tautologies hard for strong and possibly for all proof systems. We focus on a conjecture from \cite{Kra-dual} in foundations of the theory	
that there is a proof complexity generator hard for all
proof systems. This can be equivalently formulated (for p-time generators) without a reference to proof complexity notions as follows:
\begin{itemize}
\item There exists a p-time function $g$ stretching each input by one bit such that its range $rng(g)$ 
intersects all infinite $\np$ sets.
\end{itemize}
We consider several facets of this conjecture, including its links to bounded arithmetic (witnessing and independence results), to time-bounded Kolmogorov complexity, to feasible disjunction property of propositional proof systems and to complexity of proof search. 
We argue that a specific gadget generator from \cite{Kra-generator} is a good candidate for $g$.
We define a new hardness property of generators, the $\bigvee$-hardness, and shows that 
one specific gadget generator is the $\bigvee$-hardest (w.r.t. any sufficiently strong proof system). We define the class of feasibly infinite $\np$ sets  and show, assuming 
a hypothesis from circuit complexity, that the conjecture holds for all feasibly infinite
$\np$ sets. 
\end{abstract}


\section{Introduction}

A {\bf propositional proof system}
(to be abbreviated {\em pps}) in the sense of Cook and Reckhow \cite{CooRec}
is a polynomial time (p-time, shortly) binary relation $P(x,y)$ such that $\exists x P(x,y)$ defines exactly TAUT, the
set of propositional tautologies (in the DeMorgan language for definiteness).
The efficiency of a
pps $P$ is measured by the {\bf lengths-of-proofs function} $s_P$: for 
$\tau \in\mbox{TAUT}$ put
$$
s_P(\tau) \ :=\ \min \{|\pi|\ |\ P(\pi,\tau)\}\ .
$$
A pps $P$ for which $s_P(\tau)$ is bounded above by $|\tau|^c$ for some independent $c \geq 1$
is called {\bf p-bounded}. 
As pointed out by Cook and Reckhow \cite{CooRec}, 
the {\em $\np$ vs. $co\np$ problem} (asking whether the computational complexity class $\np$ is closed under
complementation) can be equivalently restated as a question whether a p-bounded pps exists. 
The existence of a p-bounded pps is thus a fundamental problem of proof complexity.

A pps $P$ is not p-bounded iff there exists an 
infinite subset $H \subseteq \mbox{TAUT}$
such that for any $c \geq 1$, for only finitely many $\tau \in H$ 
it holds that $s_P(\tau) \le |\tau|^c$.
Any such set $H$ will be said to be {\bf hard for $P$}.

 There are essentially only two classes of formulas known that make plausible candidates for being hard
for strong pps: reflection principles and $\tau$-formulas coming from proof complexity generators.
The former class is a classic topic of proof complexity and its exposition can be found in \cite[Sec.19.2]{prf}.

The latter formulas are constructed as follows. Take a function $g : \uu \rightarrow \uu$ that stretches 
all size $n$ inputs to size $m = m(n) > n$ (and hence the complement of its range $rng(g)$ is infinite) and
such that its restriction $g_n$ to $\nn$ is computed by a size $m^{O(1)}$ circuit $C_n$.
For each $b \in \mm \setminus rng(g_n)$
encode naturally (as in the proof of the $\np$-completeness of SAT)
the statement 
$$
|x|=n \rightarrow C_n(x)\neq b
$$
by a size $m^{O(1)}$ tautology $\tau(g)_b$. Function $g$
is said to be {\bf hard for $P$} iff the set $\bigcup_{n \geq 1}\{\tau(g)_b\ |\ b \in \{0,1\}^{m(n)} \setminus rng(g_n)\}$ 
is hard for $P$, and we speak of function $g$ as of a {\bf proof complexity generator} 
in this context.

We shall actually restrict ourselves here\footnote{Note 
that one can allow that the output bits of the generator $g$ are computed in non-uniform
$NTime(m^{O(1)}) \cap coNTime(m^{O(1)})$ and still
get tautologies of size polynomial in $m$ expressing that $b \notin rng(g_n)$, cf. 
Razborov \cite[Conj.2]{Raz03}, \cite[Conj.1]{Kra-finding} and \cite{Kra-di,Kra-nwg}.
There are quite a few facts known about such generators and the interested reader may start with \cite{Kra-di,Kra-nwg,Kra-finding}. }
to the rudimentary case of generators $g$ 
computed in time polynomial in $n$ (except the example of function
$\tru_{s,k}$ defined below that is computed in time polynomial in $m$) and, 
in fact, Lemma \ref{5.7.22c} shows that non-uniformity of $g$ (i.e. $g$ is computed
by a circuits that need not to come from a common algorithm)
is to some extent irrelevant.

\bigskip

The $\tau(g)_b$-formulas were defined in \cite{Kra-wphp} 
motivated by problems in bounded arithmetic
and independently (and with an apparently different motivation)
in Alekhnovich et al. \cite{ABRW}. Unfortunately the authors of \cite{ABRW} did not pursue the topic\footnote{With the sole exception of \cite{Raz03} (although published in 2015 it was written in 2002/03).} and
developing the theory was a rather lonely affair until recently. 
The theory of proof complexity generators has now a number of facets and it is
linked 
not only to bounded arithmetic and proof complexity but also to various topics in
computational complexity theory. To give the reader an idea let us mention (just by 
key words and phrases) some topics that have a non-trivial contact with the theory: 
\begin{itemize}

	\item lengths-of-proofs lower bounds, feasible interpolation, implicit proof systems, proof search,
	
	\item circuit complexity, the minimum circuit size problem, natural proofs, non-deterministic circuits, 
	
	\item bounded arithmetic, G\"{o}del's incompleteness theorem, provability of upper and lower bounds, 
		forcing with random variables, 
		
	\item Nisan-Wigderson generators, structural complexity, $\np$ search problems, Kolmogorov complexity, learning theory,
	
	\item pseudo-randomness, one-way functions, indistinguishability obfuscation.
		
	\end{itemize}
A more detailed presentation of key points of the theory and of the necessary background
requires a text of a book-length but  the interested reader may 
look at \cite[Sec.19.4-6]{prf} (or at older 
\cite[Chpts.29-30]{k2}) for an overview and further references. 
The introduction to Razborov's \cite{Raz03} is an interesting presentation of
his ideas about the topic (including a formulation of a conjecture that stimulated some
of my own work).

Be it as it may, the theory as it is now
grew out of the motivation for the formulas in \cite{Kra-wphp}:
a logic question about the provability of the dual weak PHP principle (dWPHP) for p-time functions in a weak bounded arithmetic theory $S^1_2$, cf. \cite[Problem 7.7]{Kra-wphp}.
The $\mbox{dWPHP}(f)$ says that function $f$ does not map any interval $[0,a]$ onto $[0,2a]$ (the term $2a$ can be altered
to various other values, e.g. to $a^2$ etc., without changing the logical strength of the principle over $S^1_2$). Denote the theory resulting from adding to $S^1_2$ all instances
of $\mbox{dWPHP}(f)$  for all (suitably defined) p-time functions $f$ by $\bt$. The problem
(cf. \cite[Problem 7.7]{Kra-wphp}) is:
\begin{itemize}
	\item {\em Is $\bt$ equal to $S^1_2$?  
	If not, is it at least $\Sigma^b_1$-conservative over it?
}	
\end{itemize}

This problem has a rather rich background and let me try to outline it in one paragraph.
A task inherently difficult for bounded arithmetic (and for feasible algorithms) is to
count a number of elements of a finite set. It was discovered by A.Woods \cite{Woods-phd} 
that explicit counting may be replaced in many arguments 
in combinatorics or number theory by the pigeonhole
principle PHP for bounded formulas, a statement that no bounded formula defines the graph
of a function mapping $[0,a+1]$ injectively into $[0,a]$. It is still unknown whether this
principle (denoted $\Delta_0$-PHP) is provable in bounded arithmetic (the problem is due to
MacIntyre). Then Paris, Wilkie and Woods \cite{PWW} found out that the weak PHP
(no bounded formula defines the graph of a function mapping $[0,2a]$ injectively into $[0,a]$),
denoted $\Delta_0$-WPHP, often suffices and that this principle is provable in  
bounded arithmetic (they used theory $I\Delta_0 + \Omega_1$, nowadays it is replaced
by a more convenient Buss's theory $S_2$). In a parallel development Buss 
\cite{Bus-book} defined a subtheory $S^1_2$ of $S_2$ and 
proved that functions with $\np$ graphs provably total in this
theory are exactly those that are p-time computable.
A final twist before the formulation of our problem was 
a theorem by A.Wilkie (unpublished but presented in \cite[7.3.7]{kniha}) that functions with $\np$ graphs provably total in theory $\bt$ are computable in randomized p-time.
It occurred to me that one may turn the table around and take the theory
$\bt$ as a basis for formalizing randomized algorithms 
and to establish its link with randomized p-time analogous to the link between $S^1_2$ and
deterministic p-time.
Because randomized algorithms, and probabilistic constructions and arguments more generally, 
are ubiquitous in combinatorics and complexity theory I denoted in \cite{Kra-wphp}
the theory BT for "basic theory". The link was eventually established by Je\v r\' abek 
in his PhD Thesis and in a subsequent series of papers  \cite{Jer-phd,Jer04,Jer-apc1,Jer-apc2}.
In order not to interfere with his work I decided to focus on the provability/conservativity problem above and on the related propositional logic side of things, and this lead me to proof complexity generators.

\medskip

Right from the beginning there were two working conjectures:
\begin{enumerate}

\item {\em There are generators pseudo-surjective 
for Extended Frege systems EF}, cf. \cite[Conj.7.9, Cor.7.10]{Kra-wphp},\cite[Conj.4.1,Cor.4.2]{Kra-tau}.

This conjecture is related to the provability problem mentioned above 
and  the notion of pseudo-surjectivity implies the hardness as defined earlier. 
We shall touch upon it in Section \ref{17.7.22a}, the reader can find details  
in \cite{Kra-tau,Kra-dual}.

\item {\em There is a generator hard for all proof systems}, cf. \cite[Sec.2]{Kra-dual}. 

\end{enumerate}
We shall concentrate here on the second conjecture and we shall restrict our formulation to uniform generators (i.e. computed by algorithms not just by sequences of circuits)
having the minimal required stretch $m(n) = n+1$. 
It is easy to see that
truncating any p-time generator to output-size $n+1$ preserves the hardness over any pps simulating resolution (e.g. such a truncation can be applied to generators $\tru_{s,k}$
and $U^t$ defined later). 
It also allows for a particularly simple formulation of Conjecture \ref{conj}:
by \cite[Sec.1]{Kra-dual} (or \cite[L.19.4.1]{prf})
the second conjecture can be then restated without any reference 
to proof complexity notions as follows.

\begin{conjecture}[{\cite[Sec.2]{Kra-dual}}] \label{conj}
{\ }

There exists a p-time function $g$ stretching each input by one bit such that its range $rng(g)$ 
intersects all infinite $\np$ sets. That is, the complement of $rng(g)$ is $\np$-immune.
\end{conjecture}

A fundamental question of proof complexity is, in my view, whether the hardness of proving 
a tautology can be traced back to the hardness of computing some computational task associated with the formula. A paradigm of such a reduction is the method of feasible interpolation that applies to a wide range of proof systems albeit not to strong ones (cf. \cite[Chpts.17 and 18]{prf}). One can interpret Conjecture \ref{conj} as stating
a reduction of provability hardness to computational hardness for all proof systems in
the following sense:
\begin{itemize}

\item {\em short proofs}, here witnesses to the membership in an infinite $\np$ set $A$,

\item {\em imply an upper bound on compression} for some strings in $A$, using $g$ as the decompressing algorithm. 
\end{itemize}
With a bit of imagination a direct parallel between the conjecture and 
feasible interpolation may be seen when we restrict the conjecture. 
The conjecture can be equivalently stated as asserting that all $\np$ sets disjoint
with $rng(g)$ are finite. A restriction of the conjecture may state the finiteness
just for a subclass of all $\np$ sets.
A natural restriction of Conjecture \ref{conj} in this sense, 
given a specific proof system $P$, is the restriction to $\np$ sets $A$ 
from the class of those sets for which $P$ can prove in polynomial size 
(the tautologies expressing for all lengths $n \geq 1$) that $A \cap rng(g) = \emptyset$. 
This class of $\np$ set is the resultant $Res^P_g$ of \cite{Kra-dual} and the reader
can find details there. Conjecture \ref{conj} restricted to $P$
then says that $Res^P_g$ contains only finite sets. This looks in form similar to
feasible interpolation: there we deduce feasible separability of two $\np$ sets
whose disjointness can be proved efficiently in $P$, here we deduce the finiteness of
an $\np$ set if it can be proved efficiently in $P$ that it is disjoint from a
particular $\np$ set, namely $rng(g)$.
Note also that the conjecture restricted to $P$ implies that
$P$ is not p-bounded.

\bigskip

Let us give two examples of potential generators (a third one will be discussed in 
Sec. \ref{19.7.22a}).
An illuminating example of a possibly strong generator is the {\bf truth-table function} $\tru_{s,k}$
sending a size $s$ circuit in $k$ inputs to its truth-table (a size $2^k$
string), cf. \cite{Kra-dual} or \cite[19.5]{prf}. 
Circuits of size $s$ can be coded by $10 s \log s$ bits and so to make the function
stretching we assume that $n:= 10 s \log s < m(n):=2^k$
(hence size $s$ circuits are coded by $n < m$ bits). It is computed in (uniform) time 
$O(s m) = 2^{O(k)}$, so it is p-time if $s = 2^{\Omega(k)}$.

The $\tau$-formulas determined by this generator
state circuit lower bounds for particular Boolean functions: $\tau(\tru_{s,k})_b \in \mbox{TAUT}$ iff
the function with truth-table $b$ has circuit complexity bigger than $s$.
This makes the formulas attractive but also hard to approach as we 
know very little about the size of general circuits.

It is known that the first working conjecture
above implies that the $\tau$-formulas determined by the truth table function 
$\tru_{s,k}$ (with $s = 2^{\epsilon k}$ for any $0 < \epsilon < 1$)
are hard for EF, cf. \cite{Kra-dual} or \cite[Sec.30.1]{k2}. 
On the other hand, 
unless $\nee \cap co\nee$ has size $s(k)$ circuits, the generator $\tru_{s,k}$
cannot be hard for all proof systems\footnote{But to find a pps for which it is not hard 
with any super-polynomial $s(k)$ is likely to be a hard task itself,
cf. \cite[L.29.2.2]{k2}.} 
and hence it is not a good candidate for Conjecture \ref{conj}, cf. \cite[p.198]{k2}.

\bigskip

Our second example follows \cite[Remark 6.1]{Kra-prfsearch} and concerns time-bounded Kolmogorov complexity.
Recall that the complexity measure $K^t(w)$ is the minimal size of a program that 
prints $w$ in time at most $t(|w|)$, cf. Allender \cite{All}.
The point is that a proof complexity generator
with stretch $m \geq n + \omega(1)$ produces strings $w$ of $K^t$ complexity smaller than 
$m = |w|$.
For example, if $g$ stretches $n$ bits to $m = 2n$ bits and runs in p-time $t(2n)$
then for all size $m$ strings $w \in rng(g_n)$ and $n >> 0$:
$$
K^t(w) \le  n + O(1) <  2m/3\ .
$$
In fact, as discussed in \cite[6.1]{Kra-prfsearch}, 
for a fixed polynomial time $t(n)$ sufficient for the computation of $g$ 
one can consider the universal Turing machine $U^t$ underlying the definition 
of $K^t$ as a generator itself\footnote{A similar observation was made recently 
in Ren, Santhanam and Wang \cite{RSW}.}. Then
for any pps $P$ simulating EF, if some $\tau(U^t)$-formulas
have short $P$-proofs
(e.g. by proving tautologies expressing the lower bound
$K^t(w) \geq 2m/3$),   
so do some $\tau(g)$-formulas. That is, if there is any 
$g$ computable in time $t$ and hard for $P$ then $U^t$ must be hard as well.

\bigskip

The paper is organized as follows. 
In Section \ref{17.7.22a} we consider the possibility of disproving 
(or at least of limiting possible $g$ in) Conjecture \ref{conj} by finding a feasible
way to witness that the complement of $rng(g)$ is not empty.

In Section \ref{4.7.22a} we discuss a 
new definition of hardness, the $\bigvee$-hardness, that strengthens (presumably) the hardness as defined above (but is weaker, also presumably, than 
the notion of pseudo-surjectivity mentioned earlier).
The reason for introducing the new notion is that a particular generator from the class of 
gadget generators introduced in \cite{Kra-generator}
and recalled here in Section \ref{19.7.22a}
is the $\bigvee$-hardest\footnote{Ren, Santhanam and Wang 
\cite{RSW} speak informally about the hardest proof complexity generator 
but what they define is formally an infinite family of generators.} 
among all generators but (presumably) not the hardest under  
the definition of the hardness as given above: in \cite{Kra-generator}
we used for this result the notion of iterability that 
is in strength between hardness and pseudosurjectivity mentioned in Section
\ref{17.7.22a},
as it was at hand but that is not good for Conjecture \ref{conj}.
Namely, it is known (cf. \cite{Kra-dual}) that if there is any iterable map for a given pps (containing resolution) then $\tru_{s,k}$ is iterable for it 
too and hence hard. But by the remark above $\tru_{s,k}$
is unlikely to be hard for all proof systems. 

This new notion of $\bigvee$-hardness
is equivalent to the hardness as defined above
for a class of pps satisfying the strong feasible disjunction property (Section \ref{4.7.22a}). This class
has the property that all pps {\em not in it} are automatically not p-bounded.

Section \ref{17.7.22a} is complemented
in  Section \ref{8.7.22a} where we link possible limitations to the stretch $g$ can have 
to the task of proving lower bounds on time-bounded Kolmogorov complexity. 
We argue that known results imply that these approaches are not
likely to work without proving first super-polynomial lower bounds
for (uniform and non-uniform) computations.

We also indicate in Section \ref{28.7.22a} how to modify the notion of a generator (and the conjecture and results in
Sections \ref{17.7.22a} and \ref{8.7.22a})
to address the hardness of proof search instead of lengths-of-proofs.

In Section \ref{23.8.22a} we discuss a way how to restrict Conjecture \ref{conj} and
we show, under a hypothesis, that the conjecture holds relative to all {\em feasibly infinite}
$\np$ sets: sets for which there is a p-time function picking arbitrarily large elements of the set.  
The paper is concluded by some remarks in Section \ref{19.7.22b}.

\medskip

Basic proof complexity background can be found in \cite[Chpt.1]{prf}, the topic of hard formulas
(including  a brief introduction to the theory of proof complexity generators) is in \cite[Chpt.19]{prf}. 
When we use some proof complexity notions and facts in a formal statement we define them first (and give a reference). But
we also use proof complexity background in various informal remarks and there we only
refer to the original source and/or to 
a place in \cite{prf} where it can be found.

\section{Witnessing the dWPHP} \label{17.7.22a}

The dWPHP for a function $g$ extending $n$ bits to $m = m(n)$ bits is formalized by
the formula
$$
\forall 1^{(n)} \exists y (|y| = m) \forall x (|x|=n)\ g(x) \neq y\ .
$$
Notation $\forall 1^{(n)}$ means that the universal quantifier ranges over all strings
$1\dots 1$ of any length $n$.
To witness this formula means to find a witness $y$ for the existential quantifier given
$1^{(n)}$ as input. 
This task became known recently in complexity theory as the {\em range avoidance 
problem}\footnote{That problem deals with functions computed by circuits
and the input to the task is the circuit itself; that is included in the formulation above as $g$ can have parameters (not shown in the notation).}.

Witnessing is a classic notion of proof theory\footnote{In particular,
witnessing of dWPHP is discussed in \cite[Sec.7]{Kra-dual}.} and, in particular, many fundamental results
in bounded arithmetic are formulated as follows: if a theory $T$ proves a formula of 
a certain syntactic complexity then it can be witnessed 
(i.e. its leading $\exists$ can be witnessed) by a function from 
a certain computational class $C$.
Such statements are known 
for many basic bounded arithmetic theories, many natural syntactic classes of formulas
and computational classes of functions.

Unprovability results are generally difficult and usually conditional, and we shall use one below.
But in the relativized set-up (in our situation this would mean that $g$ is given by an oracle) 
many unconditional 
unprovability results are known and they are usually derived by showing that a principle at hand cannot be witnessed
by a function in some particular class $C$ (for dWPHP see the end of this section). 

We now give an application of the conditional unprovability result of \cite{Kra-small}.
Consider theory $\tpv$ whose language has a $k$-ary function symbol $f_M$ 
attached to every p-time clocked machine $M$ with $k$ inputs, all $k \geq 1$. The symbol
$f_M$ is naturally interpreted on $\mathbf N$ by the function $M$ computes.
The axioms of $\tpv$ are all universal sentences in the language true in $\mathbf N$ under
this interpretation.

The hypothesis used in the unprovability result is this.

\medskip
\noindent
{\bf Hypothesis (H):} 

{\em There exists a constant $d \geq 1$ such that every language in $\pp$ can be decided by circuits of size $O(n^d)$:
$\pp \subseteq \mbox{Size}(n^d)$.}

\medskip

The possibility that (H) is true with $d=1$ is attributed to Kolmogorov but it is not 
a hypothesis accepted by mainstream complexity theory. However, there are no technical results
supporting the skepticism. In fact, (H) has a number of interesting consequences 
such as  $\pp \neq \np$ or $\ee \subseteq Size(2^{o(n)})$ (the latter is bad for universal derandomization but it is good for proof complexity, cf. \cite{Kra-di,Kra-small}). 

The following theorem uses 
$g := \tru_{s,k}$ with $s = 2^{\epsilon k}$ for a fixed $0 <\epsilon < 1$ for our p-time 
function. The dWPHP for this function can be expressed by formula:
\begin{equation} \label{9.4.20a}
	\forall 1^{(m)} (m=2^k > 1) \exists y \in \mm \forall x \in \nn,\ \tru_{s,k}(x) \neq y
\end{equation}
where $n = n(m) := 10 s \log s$.
We chose $m$ as the natural parameter: $m$ and $n$
are polynomially related and determine each other so this indeed expresses dWPHP. 

\begin{theorem} [{\cite{Kra-small}}] \label{22.7.22a}
{\ }

Assume hypothesis (H). Then for every $0 < \epsilon < 1$ and $s = s(k) := 2^{\epsilon k}$
the theory $\tpv$ does not prove the sentence
(\ref{9.4.20a}).
\end{theorem}

The proof of this theorem in \cite{Kra-small} goes by showing that (\ref{9.4.20a})
cannot be witnessed in a particular interactive way discussed below. However, we want
to stress that the unprovability result itself, perhaps proved from other hypotheses (or unconditionally) not using witnessing methods (but using model theory instead, for example)
implies the impossibility to witness (\ref{9.4.20a}) in a particular way.
 
To illustrate the idea simply we start by showing that (\ref{9.4.20a}) cannot 
be witnessed by  a p-time function $f$, assuming (H).
The property that $f$
witnesses (\ref{9.4.20a}) is itself a universal statement
$$
\forall 1^{(m)} \forall x (|x|=n) \ (|f(1^{(m)})| = m \wedge g(x) \neq f(1^{(m)}))
$$
and hence, if true, an axiom of $\tpv$. As this axiom easily implies (\ref{9.4.20a})
we get a contradiction with Theorem \ref{22.7.22a}. 

In fact, it is easy to see (as pointed out by one of the referees) that for any specific $0 < \epsilon < 1$
the existence of a p-time witnessing function $f$ for (\ref{9.4.20a})
with $s = 2^{\epsilon k}$
is equivalent to the existence of a language in $\ee \setminus 
Size(2^{\epsilon n})$:
the set 
$$
\{f(1^{(2^\ell)})\ |\ \ell \geq 1\}
$$ 
for any potential p-time $f$ consists of the collection of characteristic functions of a language
in $\ee$ for input lengths $\ell \geq 1$, and vice versa. 

Consider now an interactive
model of witnessing via 
Student - Teacher computation. In this model of computation (cf. \cite{KPT,KPS})
p-time student $S$, given $1^{(n)}$, produces his candidate solution
$b_1 \in \mm$. A computationally unlimited teacher $T$ either acknowledges the correctness or she produces a counter-example:
$x_1 \in \nn$ s.t. $g(x_1) = b_1$. $S$ then produces his second candidate solution $b_2$ using also $x_1$,
$T$ either accepts it or gives counter-example $x_2$ etc. The requirement is that within a given bound
$t$ on the number of rounds $S$ always succeeds. This can be written in a universal way as
\begin{equation} \label{24.8.22a}
g(x_1) \neq S(1^{(n)}) \vee
g(x_2) \neq S(1^{(n)},x_1) \vee \dots \vee
g(x_t) \neq S(1^{(n)},x_1,\dots, x_{t-1})\ .
\end{equation}
The witnessing for $\tpv$ guarantees that $t$ is a constant. 
We remark that the witnessing for $\tpv + S^1_2$ yields 
S-T protocol with polynomially many rounds $t= m^{O(1)}$; this relates to the notion of
pseudo-surjectivity mentioned in the Introduction
(the universal statement (\ref{24.8.22a}) can be represented by an infinite family of p-size tautologies and pseudo-surjectivity requires that these tautologies do not have short proofs,
cf. \cite{Kra-tau,Kra-dual} for details).

Let us state the conclusion of this discussion formally.

\begin{theorem} \label{5.7.23a}
{\ }

Assume hypothesis (H). Then dWPHP for function $\tru_{s,k}$ with parameters as in Theorem \ref{22.7.22a}
cannot be witnessed by a Student-Teacher computation with p-time Student
and constantly many rounds. 
\end{theorem}

Hence to witness the non-emptiness of the
complement of $\tru_{s,k}$ with parameters as in Theorem \ref{22.7.22a}
by a constant round S-T protocol with p-time student
would imply arbitrarily high polynomial lower bounds for circuits
computing a language in $\pp$.

Recently Ilango, Li and Williams  \cite{Ila} proved that the dWPHP for the 
circuit value function $CV$ 
(cf. Section \ref{19.7.22a}) is not provable in $\tpv$
by showing that it cannot be witnessed by an S-T computation with parameters
as in Theorem \ref{5.7.23a}, assuming  
a couple of hypotheses of a different nature: that $co\np$ is not infinitely often in 
the Arthur-Merlin class AM and a
heuristically justified conjecture in cryptography about the security of the
indistinguishability obfuscation $i{\cal O}$.
Both these hypotheses appear to be accepted by majority of experts (as oppose to hypothesis 
(H)). However,
one may wonder whether the hypothesis that the dWPHP cannot be in general witnessed
by a constant-round (or even with polynomially many rounds) S-T protocol with 
a p-time student is not more fundamental, in the sense of being closer to basic 
concepts, than the hypotheses above used to derive it.

\medskip

If we manage to extend the unprovability to theory $\tpv \cup S^1_2$ then we would rule out witnessing by S-T computation
with polynomially many rounds. Extending it further to theory $\tpv \cup T^1_2$ (or equivalently to $\tpv \cup S^2_2$)
would rule out witnessing by p-time machines accessing an $\np$ oracle.
All these statements need to be conditional as they imply (unconditionally) that $\pp$ differs from $\np$: if $\pp = \np$ then this
is implied by a true universal statement in the language of $\tpv$ (saying that a particular p-time algorithm solves SAT)
and hence all true universal closures of bounded formulas are equivalent over $\tpv$ to universal statements which are axioms of $\tpv$.

Further note that in the relativized world we have a number of unconditional results
about the impossibility to witness
dWPHP. As an example let us mention that 
we cannot witness dWPHP by a non-uniform p-time machine (i.e. using a sequence of polynomial size circuits, cf. Sipser \cite{Sip-book}) with an 
access to an $\np^{R}$ oracle where $R$ is the graph of $g$ that $g$ is not a bijection between $[0,a]$ and $[0,2a]$.
Another example is that
even if we have oracle access to $g$ and to another function $f$ we cannot witness
by a PLS problem with base data defined by p-time machines with oracle access to $f,g$
that $g$ is not a bijection between $[0,a]$ and $[0,2a]$
with $f$ being its inverse map. The interested reader can find these results (and all background)
in \cite[Secs.11.2-3]{kniha} and in references given there.

\section{Feasible disjunction property and $\bigvee$-hardness} \label{4.7.22a}

We shall propose in this section a notion of hardness that is preserved by more constructions
(and, in particularly, by the construction underlying gadget generators in Section \ref{19.7.22a})
than is the original hardness but is presumably weaker than a stronger notion of 
iterability (mentioned in the introduction) used in \cite{Kra-generator}.

\begin{definition} \label{1.7.22b}
{\ }

A function $g : \uu \rightarrow \uu$ that for any $n \geq 1$ stretches 
all size $n$ inputs to size $m:=m(n) > n$ and such that $g_n$ (the restriction of $g$ to
$\nn$) is computed by size $m^{O(1)}$ circuits
is {\bf $\bigvee$-hard} for a pps $P$ is for any $c \geq 1$,
only finitely many disjunctions
\begin{equation} \label{29.6.23c}
\tau(g_n)_{b_1} \vee \dots\vee\tau(g_n)_{b_r}\ ,
\end{equation}
with $n, r \geq 1$ and all $b_i \in \mm$, have $P$-proof of size at most $m^c$.
\end{definition}
Note that the definition can be formulated equivalently as saying that the set of 
all valid disjunctions of the form (\ref{29.6.23c}) is hard for $P$.

A pps $P$ has the {\bf feasible disjunction property} (abbreviated {\em fdp}) iff whenever a disjunction
$\alpha_0 \vee \alpha_1$ of two formulas having no atoms in common has a $P$-proof of size $s$
then one of $\alpha_i$ has a $P$-proof of size $s^{O(1)}$. The {\bf strong fdp} is defined in the same
way but the starting disjunction can have any arity $r$: ${\bigvee}_{i < r} \alpha_i$. 
The strong fdp plays a role in analysis
of a proof complexity generator in \cite{Kra-nwg}, see also \cite[Subsec.17.9.2]{prf}. It is an
open problem (\cite[Prob.17.9.1]{prf}) whether, for example, Frege or Extended Frege systems have the (strong) fdp. 
Let us note that Garl\' ik \cite{Gar} proved that 
the proof systems $R(k)$ of \cite{Kra-wphp} have no
fdp.

\begin{lemma} \label{17.7.22b}
Assume a pps $P$ has the strong fdp. Then any generator hard for $P$ is also $\bigvee$-hard for $P$.
\end{lemma}

\begin{lemma}
Assume that $g$ is a function 
stretching size $n$ inputs to size $n+1$ and such that $g_n$ (the restriction of $g$ to
$\nn$) is computed by size $n^{O(1)}$ circuits and is $\bigvee$-hard for a pps $P$.

Then for all $\delta > 0$ there is $g'$ computed by size $n^{O(1)}$ circuits and stretching
size $n$ inputs to size $n + n^{1-\delta}$ that is 
$\bigvee$-hard for $P$. 
\end{lemma}

\prf

Let $g'$ compute $g$ in parallel on $n^c$ many different inputs of size $n$:
it stretches $n^{c+1}$ bits into $n^{c+1} + n^{c}$ bits.
As the $\tau$-formulas for $g'$ are disjunctions of the $\tau$-formulas
for $g$, the lemma follows by taking $c \geq 1$ large enough.

\qed

\medskip
\noindent
{\bf A strategic choice: use $\bigvee$-hardness} 

{\em As it was pointed out in \cite{Kra-nwg}, for the purpose of proving lengths-of-proofs
lower bounds for some pps $P$ we may assume w.l.o.g. that $P$ satisfies the strong fdp: 
otherwise
it is not p-bounded and we are done. This observation, together with Lemma \ref{17.7.22b},
justifies the use of $\bigvee$-hardness rather than mere hardness.}

\medskip

The reader skeptical about the choice may interpret the statements contra-positively as sufficient conditions
refuting the strong fdp for a particular pps, cf. Lemma \ref{5.7.22b}. 
In particular, it may happen that no strong pps has the strong fdp:
but then we can celebrate as $\np \neq co\np$.

\section{The gadget generator} \label{19.7.22a}

The class of gadget generators was introduced in \cite{Kra-generator} and it is defined as follows.
Given any p-time function
$$
f\ : \ \bits^\ell \times \kk\ \rightarrow\ \bits^{k+1}
$$
define a {\bf gadget generator based on $f$} 
$$
Gad_f\ : \ \nn \rightarrow \mm
$$
where
$$
n := \ell + k (\ell +1)\ \mbox{ and }\ m := n+1
$$
as follows:
\begin{enumerate}

\item {\em The input $\overline x \in \nn$ is interpreted as $\ell + 2$ strings
$$
v, u^1, \dots, u^{\ell + 1}
$$
where $v \in \bits^\ell$ and $u^i \in \kk$ for all $i$.}

\item {\em The output $\overline y = Gad_f(\overline x)$ is the
concatenation of $\ell + 1$ strings $w^s \in \bits^{k+1}$ where we put
$$
w^s\ :=\ f(v, u^s)\ .
$$
}
\end{enumerate}
Clearly we may fix $f$ w.l.o.g. to be the 
{\bf circuit value function} $CV_{\ell,k}(v,u)$
which from a size $\ell$ description $v$ of a circuit (denoted also $v$) with $k$ inputs
and $k+1$ outputs and from $u \in \kk$ computes
the value of $v$ on $u$, an element of $\bits^{k+1}$. 

It was shown in \cite{Kra-generator} (see also \cite[L.19.4.6]{prf})
that if we replace the hardness of a generator by a stronger condition
then it suffices to consider circuits $v$ of size 
$\le k^{1 + \epsilon}$, any fixed $\epsilon > 0$. 
The proof of this fact in \cite{Kra-generator} used the notion of iterability 
mentioned earlier, as it was 
at hand. However, the same argument gives Theorem \ref{1.7.22a} using 
the presumably weaker notion
of $\bigvee$-hardness from Section \ref{4.7.22a}; the proof in \cite{Kra-generator} was only sketched so we give it here. Recall that a pps $P$ {\bf simulates} $Q$ iff
for all $\sigma \in \mbox{TAUT}$ it holds that
$s_P(\sigma) \le s_Q(\sigma)^c$.

\smallskip
\noindent
{\bf Notation:}

{\em In the rest of paper we shall ease on the notation and we
	will denote the gadget generator $Gad_f$ based on
	$f = CV_{k^2,k}$ by $\gad$ ($sq$ stands for square). }

\begin{theorem}[ess.\cite{Kra-generator}]\label{1.7.22a}
{\ }

Let $P$ be a pps simulating EF and having the following properties.
There is $c \geq 1$ such that
\begin{itemize}
	\item whenever $\sigma \in \mbox{TAUT}$
	and $\sigma'$ is obtained from $\sigma$ by substituting for some atoms 
	constants $0$ or $1$ then $s_P(\sigma') \le s_P(\sigma)^c$, and
	
	\item for all $\alpha, \beta$: $s_P(\beta) \le (s_P(\alpha) + 
	s_P(\alpha \rightarrow \beta))^c$.

\end{itemize}
Assume that there exists a 
p-time function $g : \uu \rightarrow \uu$ that stretches 
all size $n$ inputs to size $m:=m(n) > n$ and is $\bigvee$-hard for $P$. 

Then
the gadget generator based on $CV_{k^2,k}$ is $\bigvee$-hard (and hence also hard) for $P$
as well.
\end{theorem}

\prf

Assume $P$ and $g$ satisfy the hypotheses of the theorem; w.l.o.g. 
we may assume that $m(n) = n+1$.
Let $C_k$ be a canonical circuit of size polynomial in $k$ that computes $g_k$
and let $C_k$ be encoded by a string $\lceil C_k \rceil$ of size  $\ell \le k^a$, some
constant $a \geq 1$.

\medskip
\noindent
{\bf Claim 1:} {\em $\gadf$ with $f := CV_{k^a,k}$ is $\bigvee$-hard for $P$.
}

\smallskip

Note that the $\tau$ formula for $\gadf$ and $b = (b^1, \dots, b^t)\in \{0,1\}^{n+1}$ is a 
$t$-size disjunction, $t = k^a  + 1$, of $\tau$-formulas for 
$CV_{k^a,k}$ and $b^i$, $i \le t$. Substitute there for (atoms defining) 
the gadget $v := \lceil C_k \rceil$. 
Using that $EF$ has p-size proofs\footnote{When $CV_{\ell,k}$ is defined naturally by induction on the size of the circuit and the encoding $\lceil C_k \rceil$ uses $\log$-size addresses of subcircuits it would suffice to assume $P \geq R(\log)$, cf. \cite{prf} for the proof system.} 
of
$$
CV_{k^a,k}(\lceil C_k\rceil, u) = C_k(u)
$$
and $P \geq EF$, any proof of the original disjunction for $\gadf$ is turned into a polynomially longer $P$-proof of a disjuction of $\tau$-formulas for $g$, contradicting the hypothesis.

\medskip
\noindent
{\bf Claim 2:} {\em $\gad$ is $\bigvee$-hard for $P$.
}

\smallskip

Note that $\gadf$ in Claim 1 is computed in time $O(k^{2a})$ which is $\le n^{2-\delta}$
for some $\delta > 0$. Hence we may perform the same construction as in Claim 1 but using 
$\gadf$ instead of $g$ now.

\qed

Note that a circuit of size $s$ can be encoded by $10 s \log s$ bits
so $\gad$ uses as gadgets circuits of size a little bit less than quadratic.
Observe also that
$\gad$ is computed in time smaller than $n^{3/2}$.

The next statement shows that non-uniformity is irrelevant in the presence of strong fdp. 
It is proved analogously as Theorem \ref{1.7.22a}  by taking for gadgets
circuits needed to compute the generator.

\begin{lemma} \label{5.7.22c}
Assume a pps $P$ satisfies the hypotheses of Theorem \ref{1.7.22a}
and that it admits a $\bigvee$-hard proof complexity generator
computed in non-uniform p-time (i.e. by p-size circuits).
Then $\gad$ is $\bigvee$-hard for $P$. 
\end{lemma}

It is known that gadget generators (and $\gad$ in particular) are hard for many proof systems
for which we know any super-polynomial lower bound, cf. \cite{prf}. Our
{\bf working hypothesis} is that the generator $\gad$ satisfies Conjecture \ref{conj}.
But when working with the generator we encounter
the same difficulty as in the case of the truth-table generator $\tru_{s,k}$: we know nothing non-trivial
about circuits of sub-quadratic size. 
Furthermore, the experience with
lengths-of-proofs lower bounds we have so far suggests that it is instrumental to have hard examples
with some clear combinatorial structure. 
Hence to study the hardness of $\gad$ it may be advantageous to consider 
gadgets (i.e. sub-quadratic circuits) of a special form (technically that would be a substitution instance of $\gad$). 

One such specific generator was defined in \cite[pp.431-2]{prf}
and denoted $\mbox{nw}_{k,c}$ there; its gadget is essentially a slightly over-determined
system of sparse equations for a generic function $h$. 
Namely the gadget consists of:

\begin{itemize}

\item  $k+1$ sets $J_1, \dots, J_{k+1} \subseteq 
\{x_1,\dots, x_k\}$, each of size $1 \le c \le \log k$, 

\item together with $2^c$ bits defining 
truth table of a Boolean function $h$ 
with $c$ inputs. 
\end{itemize}
Given gadget $v$ and $u \in \kk$,
$f(v,u) \in \{0,1\}^{k+1}$ are the $k+1$ values $h$ computes on values that $u$ gives to variables in sets $J_1, \dots, J_{k+1}$.
This generator for one fixed, non-uniform gadget was the original suggestion for Conjecture \ref{conj} in \cite{Kra-dual} 
but the gadget generator construction allows to avoid the non-uniformity and consider generic case.

\section{Stretch and the $Kt$-complexity} \label{8.7.22a}

The main aim of proof complexity generators is to provide hard examples and for this purpose
the stretch $n+1$ of $g$ in Conjecture \ref{conj} suffices (and it yields the shortest $\tau$-formulas). 
A larger stretch is of interest in 
a connection\footnote{In fact, the need for larger stretch even in this connection seems to be eliminated by the 
	notion of iterability, cf. \cite{Kra-dual}.}
with the truth-table function $\tru_{s,k}$ discussed earlier.

We may try to limit possible stretch of generators via some considerations involving time-bounded Kolmogorov
complexity as we touched upon in the Introduction.
We shall use Levin's measure $Kt(w)$: the minimum value of
$|d| + \log t$, where program $d$ prints $w$ in time $t$, cf. Allender \cite{All}.
Its advantage over $K^t$ is that it does not require to fix the time in advance.
Although a statement like $Kt(w) \geq 2m/3$ can presumably not be
expressed by a p-size (in $m$) tautology, 
certificates for the membership in an $\np$ set $A$ such that all $w \in A$
satisfy $Kt(w) \geq 2|w|/3$ 
can be interpreted as proofs of $Kt(w) \geq 2m/3$.

Let us consider a function with an extreme stretch: $\tru_{s,k}$ with $s = 100 k$. 
This generator sends $n = 10 s \log s \le O(\log m \log\log m)$ bits 
to $m = 2^k$ bits and is computed in time $t = O(s m) < m^{3/2}$.
Hence both $K^t$ 
and $Kt$ are bounded above on $rng(\tru_{s,k}) \cap \mm$ by $O(\log m \log\log m)$.

\medskip
\noindent
{\bf Notation} (Allender \cite{All}):

{\em For any set $A \subseteq \uu$ define function
	$Kt_A : {\mathbf N}^+ \rightarrow {\mathbf N}^+$
	by
	$$
	Kt_A(m)\ :=\ \min \{Kt(w)\ |\ w \in \mm\cap A\}
	$$
	if the right-hand side is non-empty, and we leave  $Kt_A(m)$ undefined otherwise.}

\medskip

Hence we could rule out a generator with 
the extreme stretch (as in the above $\tru_{s,k}$) being hard for all 
proof systems  
if we could find an infinite $\np$ set $A$ such that
$Kt_A(m) \geq \omega(\log m \cdot \log \log m)$.
Unfortunately the next theorem suggests that this is likely not an easy task.
Following Allender \cite{All} we define an {\bf $\nee$ search problem} to be
a binary relation $R(x,y)$ such that $R$ implicitly bounds $|y|$ by $2^{O(n)}$ for $|x| = n$ and which is
decidable in time $2^{O(n)}$ (think of $y$ as an accepting computation of an $\nee$ machine on input $x$).
The search task is: given $x$, find $y$ such that $R(x,y)$, if it exists. As an example related to our situation let $A$ be an $\np$ set defined by condition 
$$
u \in A \ \mbox{ iff }\  \exists v (|v| \le |u|^c) S(u,v)
$$
with $S$ a p-time relation, and 
consider $R(x,y)$ with $y = [y_1, y_2]$ defined by:
$$
|y_1|=x \wedge |y_2| \le |y_1|^c \wedge S(y_1, y_2)\ .
$$
Note that $|y_1| = x$ expresses that the length of $y_1$ is exponential in the length
of $x$.

\begin{theorem} [{Allender \cite[Cor.7,Thm.8]{All}}] \label{29.6.23a}
	{\ }
	
	There exists an infinite $\np$ set $A$ 
	s.t. $Kt_A(m) = \omega(\log m)$ iff there exists an $\nee$ search problem
	s.t.:
	\begin{itemize}
		
		\item $\exists y R(x,y)$ is satisfied for infinitely many $x$,
		
		\item every algorithm running in time $2^{O(n)}$ solves the search problem for a finite number of
		inputs $x$ only.
	\end{itemize}
	
\end{theorem}
Hence ruling out generators with even very large stretch
means likely to prove significant computational lower bounds.
The following seems to be a natural test question.

\begin{problem} \label{28.7.22b}
	{\ }
	
	Is it true that any infinite $\np$ set $A$  contains a string $w \in A$
	with $Kt(w) < |w|$?
	That is, is it true that the set $\{w \ |\ Kt(w) \geq |w|\}$ is $\np$-immune?
	
\end{problem}

\begin{theorem} \label{1.7.23a}
	{\ }
	
	\begin{enumerate}
		
		\item If Problem \ref{28.7.22b} has the negative answer
		then the range of 
		no p-time generator $g$ stretching $n$ bits to $n + \omega(\log n)$ bits
		can intersect all infinite $\np$ sets.
		
		\item If Problem \ref{28.7.22b} has the affirmative answer
		then $\np$ is a proper subclass of $\eee$.
	\end{enumerate}
\end{theorem}

\prf

For the first part note that all strings in the range of $g_n$ ($g$ restricted to
$\nn$) have Kt-complexity at most $n + O(\log n$).

For the second part note that there is a function $g$ computable in time $2^{O(n)}$
such that the range of $g_n$ is the set of $w \in \{0,1\}^{n+1}$ with $Kt(w) \le n$.
We have that $rng(g) \in \ee$ and hence $\uu \setminus rng(g)$ is also in $\ee$ but it cannot
be - assuming the affirmative answer to the problem - in $\np$. This implies
that $\ee \not \subseteq  \np$ and hence also $\eee \not \subseteq  \np$.
As $\np \subseteq \eee$ we have
$\np \subset \eee$.

\qed

We would rather like to see the affirmative answer; not only does it have nice corollary
by the previous theorem, but it is also in the spirit 
of a potential reduction of provability hardness to computational hardness
discussed after Conjecture \ref{conj}.
Note that the problem has the affirmative answer for all $\np$ sets defined in the CSP (constraint satisfaction problem) format: if an instance $X$ of size $n$ has a solution so do instances obtained by taking $t$ disjoint copies (i.e. in disjoint sets of variables) of $X$, and these have $Kt$-complexity at most $O(n + \log t + \log tn)$ which is less than the size $tn$ of the new instance if $t > 1$ and $n >> 1$.

\bigskip

Let us consider the stretch of gadget generators. By default it was taken in the definition
to be the minimal required stretch but there are other options.
One could use as gadgets circuits that map $k$ bits to
$k'$ bits where $k' >> k$; for example, $k'=2k$ or $k'=k^2$ (allowing accordingly a bigger size
of gadgets, still polynomial in $k$). The resulting generator would send $n$ bits to approximately $(k'/k)n$ bits 
which is about $n^{1 + \epsilon}$ for some $\epsilon > 0$, for $k' = k^2$.

However, we want to be conservative with requirements on gadgets.
Note that the stretch of gadget generators can be influenced also by taking more strings $u^i$ 
in the construction of $Gad_f$ than is the minimal number
needed, i.e. more than $\ell+1$. In particular, assume we perform the construction of $Gad_f$ but taking 
$t >> \ell$ strings $u^i$ and $w^i$. We still want to maintain, as in Theorem \ref{1.7.22a}, that the generator
is the $\bigvee$-hardest generator; hence we allow only  $t$ polynomial in $k$.
Then 
$$
n := \ell + k t\ \mbox{ and }\ m := (k+1) t\ .
$$
For $\ell \le k^{O(1)}$ (as in $\gad$) and taking $t := k^c$ for very large $c \geq 1$
we can arrange that
$$
m \ \geq\ n + n^{1- \epsilon}
$$
for as small $\epsilon > 0$ as wanted. Denote the generator which extends the definition of $\gad$
in this way by $\gadc$.

\begin{lemma} \label{5.7.22b}
	{\ }
	
	Assume that there is an infinite $\np$ set $A$ such that for some $\delta > 0$:
	$$
	Kt_A(m) \geq  m - m^{1 - \delta}\ .
	$$
	Assume further that Conjecture \ref{conj} is true.
	
	Then there is a pps $P$ such that no pps $Q$ simulating $P$ has the strong fdp.
	
\end{lemma}

\prf

Choose $c \geq 1$ so large that the stretch of $\gadc$ is $n^{1-\epsilon}$,
$\epsilon = \epsilon(c)$, where
$$
m^{1-\delta} = (n + n^{1-\epsilon})^{1 -\delta} < n^{1 - \epsilon} + 2\log n
$$
for $n >> 0$ (taking $c \geq 1$ such that $0 < \epsilon(c) < \delta$ suffices).

Given an infinite $\np$ set $A$ satisfying the hypothesis 
define a pps $P$ to be, say, resolution 
but accepting also witnesses to the membership of $b \in A$ as proofs of $\tau(\gadc)_b$.
It is sound as $A$ must be disjoint from the range of $\gadc$.

If Conjecture \ref{conj} was true for some $g$ and some $Q$ simulating this $P$,
and $Q$ would satisfy the strong fdp, it would follow by 
Lemma \ref{17.7.22b} that $g$ is $\bigvee$-hard for $Q$ and hence by
Theorem \ref{1.7.22a} (modified trivially for $\gadc$)
that $\gadc$ is $\bigvee$-hard (and hence also hard) for $Q$.
That is a contradiction with how $P$ was defined.

\qed

\section{Modifications for proof search hardness} \label{28.7.22a}

Proof complexity generators, and 
Conjecture \ref{conj} in particular, aim primarily at the problem to establish 
lengths-of-proofs lower bounds. It is easy to modify the concept to aim at time complexity 
of proof search. Essentially this means to replace everywhere in the previous sections $\np$ sets by $\pp$ sets.
To give a little more detail we shall use the definition of a proof search algorithm
from \cite{Kra-prfsearch}: it is a pair $(A,P)$ such that $A$ is a deterministic algorithm that
finds for every tautology its $P$-proof.
How much time any algorithm $(A,P)$ has to use on a particular tautology is measured by the
information efficiency function $i_P : \mbox{TAUT} \rightarrow \mathbf{N}^+$; it is an inherently
algorithmic information concept.
For each pps $P$ there is a time-optimal $(A_P,P)$ 
(it has at most polynomial slow-down over any other proof search algorithm)
which is also information-optimal.
The reader can find definitions and proofs of these facts in \cite{Kra-prfsearch}.

Define a set $S \subseteq \mbox{TAUT}$ to be {\bf search-hard for $P$} iff for any $c \geq 1$
algorithm $A_P$ finds a proof of $\sigma$ in time
bounded above by $|\sigma|^c$ for finitely many formulas $\sigma \in S$ only.
Then analogously with the definition of hardness we define $g$ (in the format as in Conjecture \ref{conj}, i.e.
p-time stretching each input by one bit)
to be {\bf search-hard for $P$} iff the 
set of tautologies $\tau(g)_b$, $b \notin rng(g)$,
is search-hard for $P$.
It can be shown that
the conjecture that there is a uniform generator search-hard for all pps is then equivalent to

\begin{conjecture} [proof search version of Conjecture \ref{conj}] \label{29.6.23b}
	{\ }
	
	There exist a p-time function $g$ extending each input by one bit such that its range $rng(g)$ 
	intersects all infinite $\pp$ sets. That is, the complement of $rng(g)$ is $\pp$-immune.
\end{conjecture}

There are some more facts known about $Kt_A$ measure for sets in $\pp$ (note that Theorem 
\ref{29.6.23a} was about $\np$ sets); for example, Allender \cite[Thms.6,8]{All} or
Hirahara \cite[Thm.3.11]{Hir}.
These results seem to suggest that Conjecture \ref{29.6.23b} may not be any 
easier to prove than Conjecture \ref{conj}.

\medskip

Let us conclude this section by noticing that the fdp can be naturally modified for 
proof search as well: 
the modification requires that the
time $A_P$ needs on $\alpha_0$ or $\alpha_1$ is bounded above by a polynomial
in time it needs on $\alpha_0 \vee \alpha_1$. However, such a property implies the usual feasible interpolation
property. Namely, if $\pi$ is a $P$-proof of a disjunction 
$$
\gamma_0(x,y) \vee \gamma_1(x,z)
$$
(the disjuncts are not required to have disjoint sets of variables this time)
consider disjunction $\beta \vee (\gamma_0 \vee \gamma_1)$
where $\beta$ is a propositional sentence that is the conjunction of $0$ with all bits of $\pi$. Then $A_P$ (recall from Sec. \ref{28.7.22a} that $(A_P,P)$ is time-optimal)
when given this disjunction reads $\pi$ and hence proves  
$\gamma_0 \vee \gamma_1$ and thus also 
$\beta \vee (\gamma_0 \vee \gamma_1)$. By the search-version of fdp 
$A_P$ must find in time polynomial in $|\pi|$ a proof of $\gamma_0 \vee \gamma_1$ (as $\beta$ is false)
and thus also of any instance $\gamma_0(a,y) \vee \gamma_1(a,z)$ (this requires that $P$-proofs are closed under substitution of constants as in Theorem \ref{1.7.22a}).
By the new property again algorithm $A_P$, for each $a$  succeeds on either $\gamma_0(a,y)$ or on $\gamma_1(a,z)$
in time polynomial in $|\pi|$. That yields
feasible interpolation. 
This observation means that the proof search variant of fdp cannot hold for any strong proof systems
and is subject to same limitations as is feasible interpolation and, in particular, 
cannot hold for any strong proof systems unless some standard cryptographic assumptions fail.
The reader can find all background in \cite{prf}.

\section{Feasibly infinite $\np$ sets} \label{23.8.22a}

Two natural ways how to make 
Conjecture \ref{conj} weaker and hence more tractable are to either allow
generator $g$ from a larger class of functions than just p-time computable
or to restrict the requirement of the finiteness only to a subclass of all
$\np$ sets. 
The proof of part 2 of Theorem \ref{1.7.23a} shows
that finding $g : \nn \rightarrow \{0,1\}^{n+1}$ computable in exponential 
time would imply $\np \subset \eee$ 
(it would be that $rng(g) \in \eee$ and the argument works the same)
so such a weakening is definitely interesting
although it may not advance proof complexity. In this section we look at how to
restrict sensibly the class of $\np$ sets in the conjecture. 

We have seen one such restriction in the Introduction (classes $Res^P_g$).
There is, however, 
another natural restriction of the class of $\np$ sets in the conjecture possible. 
Take a sound theory $T$ whose language extends that of $\tpv$ consider
the class of all $\np$ sets $A$ such that the infinitude of $A$:
$$
Inf_A\ :=\ \forall x \exists y (y > x \wedge y \in A)
$$
can be proved in $T$, representing $y \in A$ by a formula in the language of $\tpv$
of the form
$$
\exists z (|z| \le |y|^c) A_0(y,z)
$$
with $c \geq 1 $ a constant and $A_0$ open and defining a p-time relation. 
Hence $Inf_A$ is an $\forall\exists$-sentence.

Knowing that a particular $T$ proves $Inf_A$ yields, in principle, 
non-trivial information about $A$.
For example, if $\tpv$ proves the sentence then by applying
Herbrand's theorem we get a p-time function $f$
witnessing it. That is, $f$ finds elements of $A$:
$$
\forall x (f(x) > x \wedge f(x) \in A)\ .
$$
We shall call sets $A$ for which such p-time function $f$ exists {\bf feasibly infinite}.
This remains true (by Buss's theorem) if $\tpv$ is augmented by $S^1_2$.
If $\tpv$ is extended by some stronger bounded arithmetic theory then $Inf_A$ will be 
witnessed by a specific $\np$ search problem attached to the theory. For example, if we add to
$\tpv$ induction axioms for $\np$ sets (theory $T^1_2$) then $Inf_A$ is witnessed by a PLS problem (by the Buss-K.theorem \cite{BK}). The reader can find the bounded arithmetic background in \cite{kniha}.

It is easy to see that Problem \ref{28.7.22b} has the affirmative answer for feasibly infinite $\np$ sets. Namely
applying function $f(x)$ to $x := 1^{(n)}$ produces $y := f(x) \in A$
with $|y| > n$ but $Kt(y) \le O(\log n)$.
For the conjecture we need to work a bit.

\begin{theorem} \label{5.7.23b}
{\ }

Assume hypothesis (H) from Section \ref{17.7.22a}. Then Conjecture \ref{conj} holds relative to the class of
feasibly infinite
$\np$ sets: there is a generator $g$ whose range intersects every feasibly infinite $\np$ set.
\end{theorem}

\prf

The proof is a special case of the construction from \cite{Kra-small}. 
We shall show that generator $\tru_{s,k}$ with $s = s(k) := 2^{k/2}$ satisfies the statement.

Let $A$ be a feasibly infinite $\np$ set as it is witnessed by a p-time function $f$.
Let $d \geq 1$ be the constant from (H) and put
$m' := m^{1/(3d)}$ where $m := |f(1^{(n)})|$ and $n >> 1$, and put also $k := \log m$.

Define the function $\hat f$ that has $m' + k$ variables
and on inputs $1^{(m')}$ and $i \in \kk$ computes the $i$-th bit of $f(1^{(n)})$; it is a p-time function. 

Take a circuit $\hat C(z,i)$ that computes $\hat f$ of size guaranteed by hypothesis (H)
and define new circuit 
$C$ by substituting $1^{(m')}$ for $z$ in $\hat C$ and leaving only
the $k$ variables for bits of $i$. Note that $C$ has size $O((m' + k)^d) < 2^{k/2}$. 
Further, by its definition, $\tru_{s,k}(C) = f(1^{(n)})$; i.e. $rng(\tru_{s,k}) \cap A \neq \emptyset$.

\qed

\begin{corollary}
	{\ }
	
Assume hypothesis (H) from Section \ref{17.7.22a}. Then 
there exists a model $\mathbf M$ of $\tpv$ in which
Conjecture \ref{conj} holds: 
there is a p-time generator $g$ such that for any standard $\np$ set $A$ (i.e. defined
without parameters from $\mathbf M$) it holds:
	$$
{\mathbf M} \ \models\	rng(g)\cap A = \emptyset\ \rightarrow
	\ \neg Inf_A \ .
	$$
\end{corollary}

\prf

Take the function $g$ from Theorem \ref{5.7.23b}. 
The statement $rng(g)\cap A = \emptyset$ is universal for any $A \in \np$,
so it is true in
the standard model $\mathbf N$ iff it is true in all models of $\tpv$. It thus
suffices to show
that $\tpv$ together with all sentences $\neg Inf_A$ for these sets $A$ is consistent.

Assume not; then the Compactness theorem and the fact that a finite number of $A_i$ 
are all disjoint from $rng(g)$ iff their union is imply that for some $\np$ set $A$
such that $rng(g)\cap A = \emptyset$ theory $\tpv$ proves $Inf_A$.
But then it is feasibly infinite and that contradicts Theorem \ref{5.7.23b}.

\qed

\section{Concluding remarks} \label{19.7.22b}

I think that it is fundamental for the development of the theory to make a progress on the original problem
of the unprovability of dWPHP for p-time functions in $S^1_2$ discussed in the Introduction. 
For a start we may try to show the unprovability in $\tpv$ (or some of its
extension as mentioned at the end of Section \ref{17.7.22a}) under a more  mainstream
hypothesis than is (H) and more theoretically fundamental than are those used in 
\cite{Ila}. Note that this presumably requires a different function than
$\tru_{s,k}$ we used in Section \ref{17.7.22a}: by remarks 
before and after Theorem \ref{22.7.22a} the unprovability of 
dWPHP for this function implies $\ee \subseteq Size(2^{o(k)})$ which contradicts
the hypothesis that $\ee \not \subseteq Size(2^{\epsilon k})$ for some $\epsilon >0$
which is - in the eyes of many complexity theorists at least - considered plausible.

\medskip

However, in my view a real progress will result only from 
unconditional results. For reasons discussed in the next-to-last paragraph of Section \ref{17.7.22a}
to have a chance to succeed we need to leave theory $\tpv$ aside and work 
with theories PV or $S^1_2$. This implies that an argument cannot rely just on witnessing theorems as they
do not change if $\tpv$ is added. The problem becomes essentially propositional and 
it is exactly this what led in \cite{Kra-tau,Kra-dual}
to the notions of freeness and pseudo-surjectivity (of generators
for EF) mentioned in Section \ref{17.7.22a}: to show that 
a p-time generator has this property is
essentially equivalent to the unprovability of dWPHP for it in PV or $S^1_2$, 
respectively (cf. \cite[Sec.6]{Kra-tau} and \cite{Kra-dual}).

\bigskip\textit{}

\noindent
{\large {\bf Acknowledgments:}} 
I thank Igor C. Oliveira (Warwick U.) and Jan Pich (Oxford U.) for discussions about the topic.
I am indebted to the two anonymous referees for their detailed comments and suggestions.


\begin{thebibliography}{99}


\bibitem {ABRW}
M.~Alekhnovich, E.~Ben-Sasson, A.~A.~Razborov,
and A.~Wigderson, Pseudorandom generators in propositional
proof complexity,
{\em SIAM J. on Computing}, {\bf 34(1)}, (2004), pp.67-88.

\bibitem {All}
E.~Allender,
Applications of Time-Bounded Kolmogorov Complexity in Complexity Theory,
in: Kolmogorov Complexity and Computational Complexity, ed.O.Watanabe,
Monographs in Theoretical Computer Science, EATCS Ser., Springer-Verlag,
(1992), pp.4-22. 

\bibitem {Bus-book}
S.~R.~Buss, {\em Bounded Arithmetic}. Naples, Bibliopolis, (1986).


\bibitem {BK}
S.~R.~Buss, and  J.~Kraj\'{\i}\v{c}ek,
An application of boolean complexity to separation problems
in bounded arithmetic,  {\em Proceedings of the London
	Mathematical Society}, {\bf 69(3)}, (1994), pp. 1-21.


\bibitem {CooRec}
S.~A.~Cook and R.~A.~Reckhow, The relative
efficiency of propositional proof systems, 
{\em J. Symbolic Logic}, {\bf 44(1)}, (1979), pp.36-50.

\bibitem {Gar}
M.~Garl\'{\i}k,  Failure of Feasible Disjunction Property for $k$-DNF Resolution 
and NP-hardness of Automating It, preprint (2020), ArXiv: 2003.10230.


\bibitem {Hir}
S.~Hirahara, Unexpected Hardness Results for Kolmogorov Complexity
Under Uniform Reductions,
in: Proc. of the 52nd Annual ACM SIGACT Symposium on Theory of Computing (STOC), June 2020, pp.1038-1051.

\bibitem {Ila}
R.~Ilango, J.~Li and R.~Williams,
Indistinguishability Obfuscation, Range Avoidance,
and Bounded Arithmetic, 
Electronic Colloquium on Computational Complexity, Report No. 38 (2023).

\bibitem {Jer-phd}
E.~Je\v{r}\'{a}bek,
{\em Weak pigeonhole principle, and randomized computation},
Ph.D. thesis, Charles University, Prague, (2005).

\bibitem {Jer04}
E.~Je\v{r}\'{a}bek,
Dual weak pigeonhole principle, Boolean complexity,
and derandomization, {\em Annals of Pure and Applied Logic},
{\bf 129}, (2004), pp.1-37.

\bibitem {Jer-apc1}
E.~Je\v{r}\'{a}bek,
Approximate counting in bounded arithmetic,
{\em J. of Symbolic Logic}, {\bf 72(3)}, (2007), pp.959-993.


\bibitem {Jer-apc2}
E.~Je\v{r}\'{a}bek,
Approximate counting by hashing in bounded arithmetic,
{\em J. of Symbolic Logic}, {\bf 7493)}, (2009), pp.829-860.


\bibitem{kniha}
J.~Kraj\'{\i}\v cek, {\em Bounded arithmetic, propositional
logic, and complexity theory},  Encyclopedia of Mathematics
and Its Applications, Vol. {\bf 60}, Cambridge University Press,
(1995).


\bibitem {Kra-wphp}
J.~Kraj\'{\i}\v{c}ek, On the weak pigeonhole principle,
{\em Fundamenta Mathematicae}, Vol.{\bf 170(1-3)}, (2001),
pp.123-140.


\bibitem {Kra-tau}
J.~Kraj\'{\i}\v{c}ek, Tautologies from pseudo-random generators,
{\em Bulletin of Symbolic Logic}, {\bf 7(2)}, (2001), pp.197-212.

\bibitem {Kra-dual}
J.~Kraj\'{\i}\v{c}ek,
Dual weak pigeonhole principle,
pseudo-surjective functions, and provability of circuit lower bounds,
{\em J. of Symbolic Logic}, {\bf 69(1)}, (2004), pp.265-286.

\bibitem {Kra-di}
J.~Kraj\'{\i}\v{c}ek,
Diagonalization in proof complexity, {\em Fundamenta Mathematicae},
{\bf 182}, (2004), pp.181-192.


\bibitem {Kra-generator}
J.~Kraj\'{\i}\v{c}ek,
A proof complexity generator, in: {\em Proc. from the 13th
Int. Congress of Logic, Methodology and Philosophy of Science (Beijing, August 2007)}, King's College Publications, London, ser. Studies in Logic and the Foundations of Mathematics. Eds. C.Glymour, W.Wang, and D.Westerstahl, (2009), pp.185-190.

\bibitem {k2}
J.~Kraj\'{\i}\v{c}ek,
{\em Forcing with random variables and proof complexity}, 
London Mathematical Society Lecture Note Series, No. {\bf 382}, 
Cambridge University Press, (2011).


\bibitem {Kra-nwg}
J.~Kraj\'{\i}\v{c}ek,
On the proof complexity of the Nisan-Wigderson generator
based on a hard $NP \cap coNP$ function,
{\em J. of Mathematical Logic}, {\bf 11(1)}, (2011), pp.11-27.

\bibitem {Kra-finding}
J.~Kraj\'{\i}\v{c}ek,
On the computational complexity of finding hard tautologies,
{\em Bulletin of the London Mathematical Society}, {\bf 46(1)}, (2014), pp.111-125.



\bibitem{prf}
J.~Kraj\'{\i}\v{c}ek, \textit{Proof complexity}, 
Encyclopedia of Mathematics and Its Applications, Vol. \textbf{170}, Cambridge University Press,
(2019).

\bibitem {Kra-small}
J.~Kraj\'{\i}\v{c}ek, 
Small circuits and dual weak PHP in the universal theory of p-time algorithms,
{\em ACM Transactions on Computational Logic}, 22, 2, Article 11 (May 2021).

\bibitem {Kra-prfsearch}
J.~Kraj\'{\i}\v{c}ek, 
Information in propositional proofs and algorithmic proof search,
{\em J. of Symbolic Logic}, vol.87, nb.2, (June 2022), pp.852-869.

\bibitem {KPS}
J.~Kraj\'\i\v{c}ek, P.~Pudl\'ak, and
J.~Sgall, Interactive
Computations of Optimal Solutions, in: B. Rovan (ed.):
{\em Mathematical
	Foundations of Computer Science}
(B. Bystrica, August '90), Lecture
Notes in Computer Science {\bf 452},
Springer-Verlag, (1990), pp. 48-60.


\bibitem {KPT}
J.~Kraj\'\i\v{c}ek, P.~Pudl\'ak and G.~Takeuti,
Bounded arithmetic and the polynomial hierarchy,
{\em Annals of Pure and Applied Logic}, {\bf 52},
(1991), pp.143--153.

\bibitem {PWW}
J.~Paris, A.~J.~Wilkie and A.~Woods,
Provability of the Pigeonhole Principle and the Existence of Infinitely Many Primes,
{\em J. of Symbolic Logic}, {\bf  53(4)}, (1988), pp.1235-1244.


\bibitem {Raz03}
A.~A.~Razborov,
Pseudorandom generators hard for $k$-DNF resolution
polynomial calculus resolution,
{\em Annals of Mathematics}, {\bf 181(2)}, (2015), pp.415-472.

\bibitem {RSW}
H.Ren, R.Santhanam and Z.Wang, 
On the Range Avoidance Problem for Circuits, ECCC Report nb.48, (2022).

\bibitem {Sip-book}
M.~Sipser, {\em Introduction to the Theory of Computation}, Cengage Learning (3rd ed.), 2005.

\bibitem {Woods-phd}
A.~Woods,
{\em Some problems in logic and number theory, and their connections},
PhD Thesis, U. of Manchester, (1981).


\end{thebibliography}
\end{document}